\documentclass[prb,twocolumn,showpacs]{revtex4}
\usepackage{epsfig}

\begin{document}

\title{Self-interaction effects in (Ga,Mn)As and (Ga,Mn)N} 

\author{Alessio Filippetti}
\affiliation{SLACS and Dipartimento
di Fisica, Universit\`a di Cagliari, I-09042 Monserrato (Ca), Italy}
\author{Nicola A. Spaldin}
\affiliation{Materials Department, University of California,
Santa Barbara, CA 93106-5050}
\author{Stefano Sanvito}
\email{sanvitos@tcd.ie} 
\affiliation{Physics Department, Trinity College, Dublin 2, Ireland}

\date{\today}

\begin{abstract}
The electronic structures of Mn-doped zincblende GaAs and wurtzite GaN are  
calculated using both standard local-density functional theory (LSDA), and 
a novel pseudopotential self-interaction-corrected approach (pseudo-SIC), 
able to account for the effects of strong correlation. We find that, as
expected, the self-interaction is not strong in (Ga,Mn)As, because the 
Fermi energy is crossed by weakly correlated As p - Mn d hybridized bands 
and the Mn 3d character is distributed through the whole valence band 
manifold. This result validates the extensive literature of LSDA studies 
on (Ga,Mn)As, including the conclusion that the ferromagnetism is 
hole-mediated. In contrast, the LSDA gives a qualitatively incorrect 
band structure for (Ga,Mn)N, which is characterized by localized Mn 3d bands 
with very strong self-interaction. Our pseudo-SIC calculations show a highly 
confined hole just above the Fermi energy in the majority 
band manifold. Such a band arrangement is consistent with (although by no 
means conclusive evidence for) a recent suggestion\cite{ZR} that formation 
of Zhang-Rice magnetic polarons is responsible for hole-mediated 
ferromagnetism in (Ga,Mn)N.
 
\end{abstract}

\pacs{Valid Pacs}  
\maketitle

\section{Introduction}

The discovery of ferromagnetism in Mn-doped GaAs and InAs \cite{ohno1} 
has changed conventional thinking about electronics. 
In fact these diluted magnetic semiconductors (DMSs) combine the functionalities of 
semiconductors with those of magnetic materials, paving the way for a complete 
integration of data storage and logic on the same electronic component \cite{wolf}. 
Unfortunately the most widely studied DMS, (Ga,Mn)As, shows a Curie temperature 
($T_C$) far below room temperature and therefore appears to be unsuitable
for commercial applications. This has led to a considerable world-wide effort in 
searching for new DMSs with higher $T_C$s. 

Part of this research is guided by a pioneering theoretical work by Dietl and 
collaborators, who predicted remarkably high $T_C$s for wide-gap magnetic 
semiconductors at moderate Mn doping \cite{dietl}. Their model is based on the 
Zener mechanism for ferromagnetism, according to which a conduction electron 
can hop across two magnetic atoms through a double exchange of place with the 
valence electrons of the bridging anions. Spin-conservation implies that this 
conduction can only occur between spin-aligned magnetic centers, thus a 
ferromagnetic ordering of the Mn ions is expected. In the model, $T_C$ is crucially 
dependent on hole and Mn concentration, and on the strength of their mutual coupling. 
High $T_C$ ferromagnetism in wide-gap DMSs has recently been confirmed by 
several groups who reported the existence of above-room-temperature 
ferromagnetism in Ga$_{1-x}$Mn$_x$N \cite{gamnn1,gamnn2,gamnn3,gamnn4}. 
However, in contrast to ``conventional'' As-based DMSs, where the existence of
double-exchange induced ferromagnetism is clear, the origin and the nature
of the observed magnetism in Ga$_{1-x}$Mn$_x$N are controversial. 
On one hand, both $p$- and $n$-doped samples are, apparently, 
ferromagnetic, thus indicating that hole-mediated double-exchange may not be, in fact, 
the driving mechanism. More importantly, recent magnetic circular dichroism 
(MCD) measurements \cite{ando} on high $T_C$ samples show that the MCD signal 
in the vicinity of the GaN band-gap comes from a paramagnetic phase. 
This suggests the possibility of two phases in (Ga,Mn)N: a paramagnetic phase,
attributed to substitutional Mn at the Ga site, and dominated by $s,p$-$d$ 
interactions between Mn and the host GaN, and an unidentified ferromagnetic
phase, not detected by MCD, which might be due to small particles that are 
invisible to diffraction methods\cite{ando}. 

In summary, there are sound indications that (Ga,Mn)As and (Ga,Mn)N may 
not be, in fact, very similar as far as magnetic ordering is concerned.
Thus a careful examination of their microscopic properties is necessary  
in order to give a correct interpretation of the experimental results.

With this goal in mind, in this paper we investigate 
the electronic structures of zincblende (Ga,Mn)As and wurtzite (Ga,Mn)N
using first-principles density functional theory (DFT) calculations. 
Band energy calculations for these compounds obtained within the familiar 
local-spin density functional theory (LSDA) are already present in the
literature.\cite{mer,me,kalotov,fong} Here we present the results 
obtained with our recently developed pseudopotential self-interaction 
free approach (pseudo-SIC).\cite{fh_sic}

It is known that the band energies calculated within LSDA have systematic 
errors caused by the presence in the one-electron Kohn-Sham (KS) potential of 
its self-interaction (SI)\cite{si} (that is the interaction of an electron
with the potential generated by the electron itself). The SI is especially 
relevant in ionic and mixed ionic-covalent materials. Indeed
the poor LSDA description of the electronic properties of II-VI 
semiconductors \cite{vogel}, transition metal oxides \cite{tmo}, and 
magnetic perovskites \cite{pero} is a consequence of the SI \cite{fh_sic}. 
In general, any compound characterized 
by localized electron charges is, in principle, affected by a 
non-vanishing, unphysical, SI contribution.

Regarding the compounds investigated in this paper, it is known that the 
LSDA band structure of bulk GaN shows severe discrepancies with photoemission 
results (for example the fundamental energy gap is severely 
underestimated and the position of the $d$ bands is incorrect). Furthermore, 
the LSDA is generally inaccurate in describing the electronic properties of 
oxides of manganese, and in particular fails to give the correct 
position of the Mn d states with respect to the oxygen-derived valence manifold. 

Since the band gap of the host material and the position of the transition metal 
d states are crucial in the description of these diluted magnetic semiconductors,
there is the necessity to confront the LSDA description with alternative 
approaches. Our pseudo-SIC is ideally suited for this task since
it repairs to a large extent the LSDA failures 
and provides a reliable description of the electronic properties for a 
large range of materials, including GaN and Mn-based perovskites. 
In addition, and in contrast with other beyond-LSDA approaches, our method 
preserves the conceptual simplicity and the computational feasibility of the
LSDA, and thus the possibility to approach large-sized systems such as the
diluted magnetic semiconductors.

In this paper we determine the extent to which correction of the SI changes 
the electronic properties and the chemical pictures of wurtzite (Ga,Mn)N 
and zincblende (Ga,Mn)As. We find that in (Ga,Mn)As the localization of 
the Mn 3d electrons is weak because both on-site and off-site hybridizations 
occur, and, as a consequence, the SI is not strong. Therefore the LSDA already 
provides a good description of (Ga,Mn)As and the pseudo-SIC does not make 
large qualitative changes. In (Ga,Mn)N, however, the SI is much stronger, 
due to the large N electronegativity, and consequent higher
ionicity of GaN. 
Therefore the LSDA description is inadequate and the pseudo-SIC
gives qualitatively different results. The difference in localization between
the two compounds in turn has a profound effect on their electronic properties.

This paper is organized as follows: in section \ref{method} we describe the
methodology and the technicalities used for the calculations. Sections \ref{gamnas} 
and \ref{gamnn} present the results for (Ga,Mn)As, and (Ga,Mn)N, 
respectively. Finally, in sections \ref{disc} and \ref{concl} we 
present our conclusions.

\section{Methodology}
\label{method}

In this paper we use our recently developed self-interaction free local-density scheme 
\cite{fh_sic} based on ultrasoft pseudopotentials\cite{uspp} and a plane wave basis. 
The scheme is described in detail in Ref. \cite{fh_sic}; here we summarize the key
points.
The Kohn-Sham equations are modified in order to effectively extract 
from the one-electron potential the spurious self-interaction contribution:

\begin{eqnarray}
\left[-\nabla^2 + \hat{V}_{\mbox{pp}} + \hat{V}_{\mbox{hxc}}^{\sigma}
- \hat{V}_{\mbox{sic}}^{\sigma}\right]|\psi^{\sigma}_{n{\bf k}}\rangle=
\epsilon^{\sigma}_{n{\bf k}}\,|\psi^{\sigma}_{n{\bf k}}\rangle,
\label{ks-sic}
\end{eqnarray}
where $\hat{V}_{\mbox{pp}}$ is the usual ion-core pseudopotential projector, 
$V_{\mbox{hxc}}^{\sigma}$ the LSDA-KS potential, and $\hat{V}_{\mbox{sic}}^{\sigma}$ 
the SIC term. The latter is cast as a nonlocal Kleinman-Bylander-type 
pseudopotential projector:
\begin{eqnarray}
\hat{V}_{\mbox{sic}}^{\sigma}\,=\,\sum_i\>
{|\,V_{\mbox{hxc}}^{\sigma,i}\>\phi_i\,\rangle\> p_i^{\sigma}\>\langle\,
\phi_i\>V_{\mbox{hxc}}^{\sigma,i}\,|
\over \langle\,\phi_i\,|\,V_{\mbox{hxc}}^{\sigma,i}\,|\,\phi_i\,\rangle}.
\label{sic1}
\end{eqnarray}
Here $\phi_i({\bf r})$ is the pseudowavefunction, 
$V_{\mbox{hxc}}^{\sigma,i}[\phi_i^2({\bf r})]$ the SI potential of the $i^{th}$ 
atomic orbital ($i$ runs over angular quantum numbers and atomic positions, 
$\sigma$ over the spin), and $p_i^{\sigma}$ is the corresponding occupation number 
calculated self-consistently by projecting the Bloch state manifold onto the basis of 
atomic pseudowavefunctions. This procedure is no more computationally demanding than 
the usual LSDA, and, in contrast with other SIC schemes \cite{vogel}, 
it can be applied to both metals and insulators, a feature which is crucial for the 
present application.

In earlier work our method was applied to a number of different 
systems including III-V and II-VI ionic semiconductors (e.g. GaN and ZnO),\cite{fh_sic} 
transition metal monoxides (MnO, NiO)\cite{fh_sic} and perovskites 
(CaMnO$_3$, BaTiO$_3$, YMnO$_3$)\cite{bec_sic}. In all cases the pseudo-SIC
obtained both the correct physical structures, and band structures 
in excellent agreement with photoemission experiments. 
This series of successes lends strong support to the credibility of the 
present results.

We use ultrasoft pseudopotentials to allow moderate cut-off energies 
(35 Ryd and 40 Ryd for (Ga,Mn)As and (Ga,Mn)N, respectively). For the
pseudo-SIC calculations the Ga 3d electrons are treated as core states 
(this choice is fully justified in Ref. \cite{fh_sic}.)
For zincblende (Ga,Mn)As and wurtzite (Ga,Mn)N 16-atom and 32-atom supercells 
are used respectively, in each case with one Mn ion substituting for one Ga ion,
corresponding to doping levels of $\sim$ 12\% and $\sim$ 6\%.
The lattice parameters are fixed to the values calculated for the host materials 
(a = 10.68 bohr for GaAs, a = 6.03 bohr, c= 9.802 bohr, u= 0.377 for GaN), 
which are in good agreement with the experimental data.

\section{Zincblende (Ga,Mn)As}
\label{gamnas}

\subsection{LSDA}

\begin{figure}
\epsfxsize=6.5cm
\centerline{\epsffile{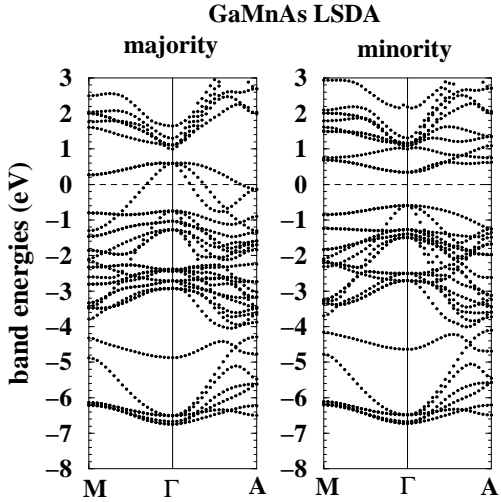}}
\caption{Band structure of zincblende (Ga,Mn)As calculated within 
LSDA for majority (left) and minority (right) spins.} 
\label{mnas_band}
\end{figure}

The calculated LSDA band structure of (Ga,Mn)As (in figure 
\ref{mnas_band}) is that of an half-metal with metallicity in the
majority spin band. Here the most important feature for the
electronic and magnetic behavior is given by the three 
bands crossing the Fermi energy in the majority manifold.
They are rather dispersed and span $\sim$2 eV between $\Gamma$ and M.
An orbital decomposition of these bands shows that they consist of Mn
d $t_2$ character, as well as Mn and As $p$, with significant
contributions from both nearest and next-nearest neighbor As ions.
This indicates that part of the charge is extended beyond the 
Mn-centered tetrahedron. The next set of bands lower in energy
corresponds to the valence band top of bulk GaAs at the
$\Gamma$ point ($E= -0.73$eV) but mixes in Mn character 
elsewhere in the Brillouin Zone, as does the doublet below it.
An orbital decomposition shows that the Mn d character
is not restricted to a single set of bands close to the Fermi 
level, but instead is widely distributed through the GaAs sp valence 
manifold.

In the upper part of Table \ref{tab_gamnas_lsd1} we report the occupation numbers 
of individual orbitals. The total magnetization per cell is $\sim$ 4 $\mu_B$, 
almost entirely due to the Mn (3.72~$\mu_B$). The Mn d electron 
charge is 5.48, of which 4.61 and 0.87 comes from spin-up and spin-down 
electrons, respectively.

\begin{table}
\caption{Occupation numbers for selected orbitals of (Ga,Mn)As calculated 
within LSDA (top) and pseudo-SIC (bottom). The arrows indicate spin-up (majority)
and spin-down (minority) components, respectively. As$_{\mbox{1n}}$ indicates the
As atoms adjacent to the Mn ion.}
\label{tab_gamnas_lsd1}
\centering\begin{tabular}{ccccccc} \hline\hline
                     & \multicolumn{2}{c} {Ga}  &  \multicolumn{2}{c} {As$_{\mbox{1n}}$} & \multicolumn{2}{c} {Mn} \\
\hline
                     & $\uparrow$ & $\downarrow$ & $\uparrow$ & $\downarrow$ & $\uparrow$ & $\downarrow$   \\
\hline
   s                 & 0.52     &  0.52     &   0.70     & 0.70          &  0.31 &  0.26   \\
   p$_x$             & 0.31     &  0.31     &   0.59     & 0.60          &  0.24 &  0.17    \\
   d t$_{2}$        &          &           &            &               &  0.89 &  0.25   \\
   d e           &          &           &            &               &  0.97 &  0.06   \\ \hline\hline
   s                 & 0.48     &  0.48     &   0.70     & 0.70          &  0.29 &  0.25   \\
   p$_x$             & 0.30     &  0.30     &   0.59     & 0.65          &  0.23 &  0.17    \\
   d t$_{2}$            &          &           &            &           &  0.97 &  0.12   \\
   d e           &          &           &            &           &  0.99 &
0.03   \\ \hline\hline
\end{tabular}
\end{table}

In conclusion, according to the LSDA description (and in agreement with previous 
calculations \cite{mer,me}) (Ga,Mn)As is dominated by Mn d - As p hybridization 
rather than on-site electron correlation. However, since the overestimation of 
p-d hybridization and the suppression of on-site Coulomb energy are general 
tendencies of the LSDA, a comparison with the pseudo-SIC is necessary in order
to assess the reliability of this picture. 

\subsection{pseudo-SIC}

\begin{figure}
\epsfxsize=6.5cm
\centerline{\epsffile{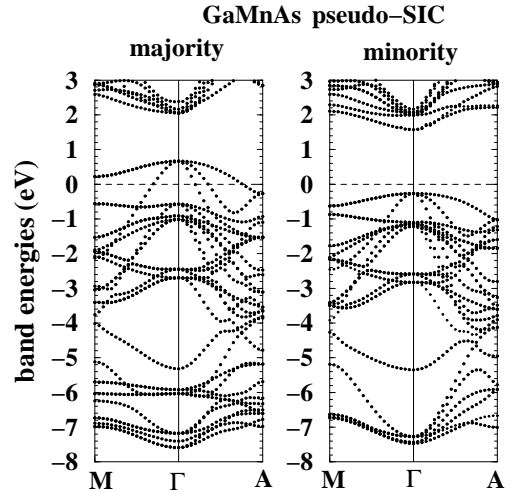}}
\caption{Band structure of zincblende (Ga,Mn)As calculated within 
pseudo-SIC for majority (left) 
and minority (right) spins.} 
\label{mnas_bandsic}
\end{figure}

In Figure \ref{mnas_bandsic} we show the band structure calculated within 
pseudo-SIC. It is immediately apparent that the main features are not 
changed significantly with respect to the LSDA, apart for the expected 
increase of the fundamental energy gap of the bulk GaAs due to the downward 
shift of the occupied valence band manifold with respect to the empty bands.
Also the band manifold occupies an energy region $\sim$ 1 eV
wider than that in the LSDA result.
Interestingly, the three majority, half-occupied bands have an even larger 
dispersion than in LSDA, but the shape of the bands is very similar.

In the lower part of Table \ref{tab_gamnas_lsd1} we report the orbital occupations calculated 
within pseudo-SIC. The total magnetization is 4 $\mu_B$, as in LSDA.
However, as a consequence of a larger Mn d spin splitting, the Mn magnetic 
moment (4.58 $\mu_B$) is larger than in LSDA, but it is compensated by a
non discardable spin-polarization on the As (0.17~$\mu_B$) which is antiparallel 
to the Mn moment. (Table \ref{tab_gamnas_lsd1} reports the values for a single
orbital (p$_x$) which has a magnetization of -0.06 $\mu_B$ in pseudo-SIC. The total
magnetization for p$_x$ + p$_y$ + p$_z$ is -0.17~$\mu_B$.)

In conclusion, the pseudo-SIC result is a sound confirmation that the p-d 
hybridization and the weakly-correlated character of the Mn~d electrons is 
in fact a reliable prediction of the LSDA, and validates the extensive 
earlier computational literature on (Ga,Mn)As. (for a review see Ref.\cite{mer}).

\section{Wurtzite (Ga,Mn)N}
\label{gamnn}

\subsection{LSDA}

\begin{figure}
\epsfxsize=6.5cm
\centerline{\epsffile{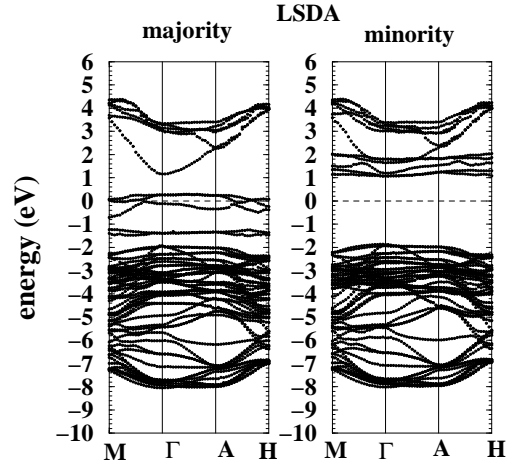}}
\caption{Band structure of wurtzite (Ga,Mn)N calculated within LSDA 
for majority (left) and minority (right) spins.} 
\label{mnn_band}
\end{figure}

\begin{figure}
\epsfxsize=6.5cm
\centerline{\epsffile{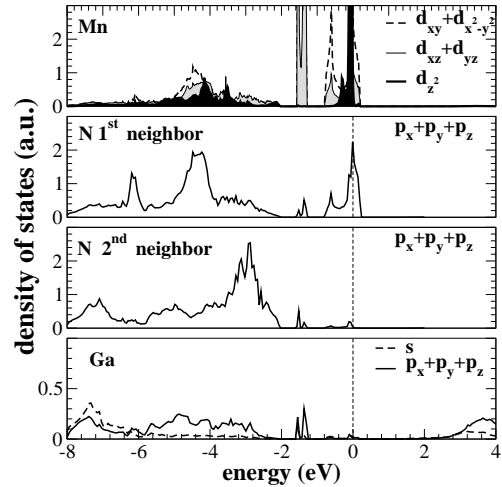}}
\caption{Orbital-resolved density of states of wurtzite (Ga,Mn)N calculated 
within LSDA. Only majority-spin orbitals are shown. For Mn, light-shaded
and dark-shaded areas represent the contributions from doublet (d$_{xz}$,d$_{yz}$) 
and singlet d$_{z^2}$, respectively. }
\label{mnn_dos}
\end{figure}

In the hexagonal crystal symmetry of the wurtzite structure, 
the d~orbitals are split into two 
doublets ((d$_{xy}$, d$_{x^2-y^2}$), and (d$_{xz}$, d$_{yz}$)) and 
one singlet (d$_z^2$), where $x$ and $y$ lie in the hexagonal plane and $z$ is 
parallel to the c axis. Similarly, the p orbitals are split into a doublet 
(p$_x$, p$_y$) and a singlet (p$_z$). In Figure \ref{mnn_band} we report 
the band structure of wurtzite (Ga,Mn)N calculated within the LSDA. Like 
(Ga,Mn)As, this material is half-metallic and has three partially filled 
bands close to the Fermi energy. However this three-band manifold (consisting
of a doublet composed of Mn d$_{xy}$-d$_{x^2-y^2}$  and the
p$_x$-p$_y$ orbitals of the three planar N, and a singlet derived from
Mn d$_{z^2}$-p$_z$ and on-top N p$_z$) is much narrower in (Ga,Mn)N than in 
(Ga,Mn)As. The doublet below these states ($E$~=~--1.39 eV) 
at $\Gamma$ is also 
a p-d combination (of Mn d$_{xy}$-d$_{x^2-y^2}$, and N p$_x$) with a strongly
localized electron charge confined almost entirely to orbitals internal to the 
Mn-centered tetrahedron. 
The small band dispersion (less than 1 eV) is evidence of the strongly-correlated 
character. The same features can be appreciated in the orbital-resolved density 
of states, reported in Figure \ref{mnn_dos} (only majority spin components are shown).
The hybridization between Mn d and the p orbitals of the four 
surrounding N is clearly visible, while almost no contribution comes 
from second neighbor N.  

The orbital occupation numbers for (Ga,Mn)N are given in the upper part of 
Table \ref{tab_gamnn_lsd}. The Mn magnetic moment is 3.63 $\mu_B$ with a 
total Mn d electron charge of 5.3. Notice how the much stronger electronegativity 
of N with respect to As is reflected in the comparison between the orbital 
occupations in (Ga,Mn)As (Table \ref{tab_gamnas_lsd1}) where roughly 60\% 
and 30\% of the s, p electron charge is distributed to As and Ga,
and in (Ga,Mn)N, where the proportion is 70\% to N and 20\% to Ga.

These band structures agree qualitatively with earlier LSDA 
calculations\cite{leeor,mark}, although the exact position of the Mn 
d levels differs according to the details of the pseudopotentials 
or the lattice parameters used in the calculations.

\subsection{pseudo-SIC}

\begin{figure}
\epsfxsize=6.5cm
\centerline{\epsffile{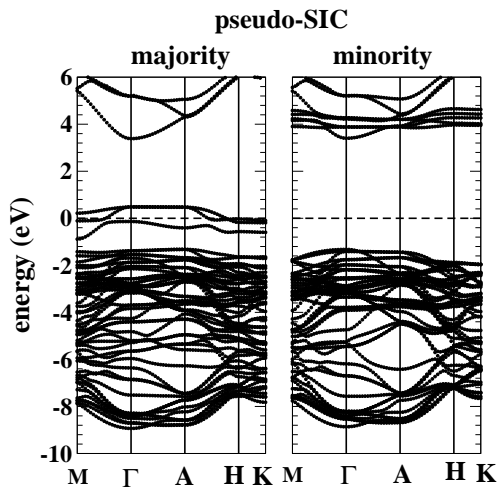}}
\caption{Band structure of wurtzite (Ga,Mn)N calculated within pseudo-SIC for 
majority (left) and minority (right) spins.} 
\label{mnn_bandsic}
\end{figure}

\begin{figure}
\epsfxsize=6.5cm
\centerline{\epsffile{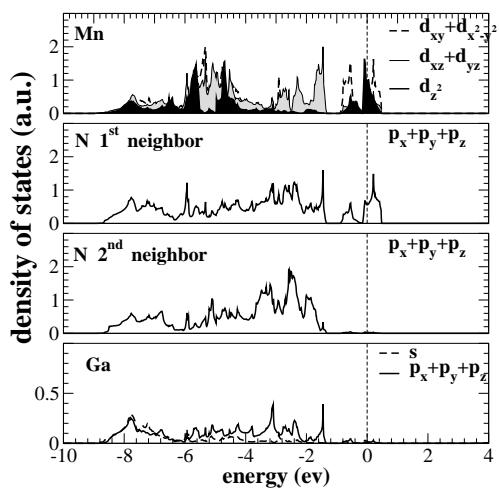}}
\caption{Orbital-resolved density of states of wurtzite (Ga,Mn)N calculated 
within pseudo-SIC. Only majority-spin orbitals are shown.}
\label{mnn_dos_sic}
\end{figure}

Strong electron localization makes the LSDA results for (Ga,Mn)N 
potentially affected by a large self-interaction. Our calculations
indeed confirm these expectations. The pseudo-SIC band 
structure and the orbital-resolved density of states are shown  
in Figures \ref{mnn_bandsic} and \ref{mnn_dos_sic}),
respectively.  

Due to the self-interaction, the fully-occupied d$_{xz}$, and d$_{yz}$ bands
are pushed down in energy, well below the p-like valence band top of the
host material. Around the Fermi energy two electrons
occupy the other three majority Mn d bands, similarly to what we have 
seen within LSDA. However now these three bands are closer to 
the top of the p band manifold, and, due to the large increase of the 
GaN energy gap, much farther from the bottom of the s-like conduction bands. 
 
In the lower part of Table \ref{tab_gamnn_lsd} the occupation 
numbers are shown. The local magnetic moments are 4.18~$\mu_B$ for Mn, 
0.06~$\mu_B$ for the on-top N, and --0.10~$\mu_B$ for each of the three 
planar N. This means that the coupling between the on-top N p$_z$ orbital
and the majority d$_{z^2}$ orbital is ferromagnetic, whereas the coupling between 
planar N $p_x$ and p$_y$ orbitals with d$_{xy}$ and d$_{x^2-y^2}$ orbitals
is antiparralel, thus contributing negatively to the total magnetic moment
M=4 $\mu_B$.

This remarkable increase of Mn magnetic moment with respect to the
LSDA values is a typical self-interaction correction effect: 
the spatial localization is increased and the hybridization
reduced.
 
Finally, notice that these results are not very sensitive to the Mn 
concentration: band energy calculations of wurtzite (Ga,Mn)N with 12\% 
Mn doping within both LSDA and pseudo-SIC shows no relevant changes 
with respect to the corresponding calculations at 6\% concentration.
    
\section{Discussion}
\label{disc}

The chemical picture of (Ga,Mn)N obtained within the pseudo-SIC (which we 
believe to be the most accurate representation) is markedly different 
both from the LSDA description of (Ga,Mn)N, and also the behavior 
of (Ga,Mn)As.
The difference between (Ga,Mn)N and (Ga,Mn)As can be traced
back to the large electronegativity of N, which in turn favors ionicity
and charge transfer over hybridization. This aspect, already evident
within LSDA, is further emphasized within pseudo-SIC, due to its
tendency to favor electronic configurations with completely filled 
or empty orbitals.\cite{note}

The differences are so large that they will certainly lead to different 
charge mobilities in the two materials, and affect the nature of the 
magnetic ordering as well.
Indeed the band alignment obtained within pseudo-SIC for (Ga,Mn)N may 
not be appropriate for validating the usual Zener picture for 
ferromagnetism. In fact, according to our calculation, 
there are no free holes in the GaN valence band. Instead, the holes 
are localized near the Mn impurity, and occupy a triplet of 1~eV wide 
bands around the Fermi energy. This could lead to two possible scenarios. 
On the one hand substitutional Mn in GaN could be responsible 
only for the paramagnetic component of the magnetization as indicated
by MCD measurements\cite{ando}. On the other hand this could be indicative of the 
formation of a Zhang-Rice polaron in (Ga,Mn)N as has been recently 
suggested \cite{ZR}. According to this model, Mn doping in GaN behaves
as an effective mass acceptor, with a configuration of d$^5$ plus a 
hole localized on the magnetic ion. This local singlet could move 
through the sublattice of Mn$^{2+}$ ions and cause
ferromagnetic ordering (still mediated by a double-exchange mechanism) 
at much higher T$_c$ then that of the usual Zener ``free hole''
double exchange. Our calculations show a magnetic moment for Mn of 
4.18~$\mu_B$, which is suggestive of a d$^4$ configuration. However the 
presence of very flat, partially occupied bands with d character allow 
us to re-interpret our results in terms of a Mn d$^5$ configuration plus a 
tightly bound hole of d-character. These are the conditions for the 
formation of a Zhang-Rice polaron. To further assess the validity of 
these hypotheses, 
larger-sized calculations, capable of describing the properties of 
these compounds in different magnetic orderings (e.g. paramagnetic
and antiferromagnetic) will be necessary in the future.

\begin{table}
\caption{Occupation numbers for selected orbitals of wurtzite (Ga,Mn)N 
calculated within LSDA (top) and pseudo-SIC (bottom). Only the 4 nitrogens 
surrounding Mn are considered: N$_{\mbox{top}}$ is that on-top of Mn, 
N$_{\mbox{base}}$ s one of the three N at the vertices of the triangular base. 
The arrows indicate spin-up (majority) and spin-down (minority) contributions, 
respectively.
\label{tab_gamnn_lsd}}
\centering\begin{tabular}{ccccccccc} \hline\hline
        & \multicolumn{2}{c} {Ga}  &  \multicolumn{2}{c}{N$_{\mbox{top}}$} & \multicolumn{2}{c}{N$_{\mbox{base}}$} & 
\multicolumn{2}{c} {Mn}\\
\hline 
                     & $\uparrow$ & $\downarrow$ & $\uparrow$ & $\downarrow$ & $\uparrow$ & $\downarrow$  
& $\uparrow$ & $\downarrow$  \\
\hline
   p$_x$             & 0.24     &  0.24     &   0.73     & 0.71  & 0.70    &   0.70      &  0.20 &  0.15    \\
   p$_z$             & 0.23     &  0.23     &   0.72     & 0.71  & 0.74    &   0.71      &  0.20 &  0.15   \\  
   d$_{xy}$          &          &           &            &       &         &             &  0.85 &  0.20   \\
   d$_{xz}$          &          &           &            &       &         &             &  0.93 &  0.14   \\
   d$_{z^2}$         &          &           &            &       &         &             &  0.82 &  0.24   \\ \hline\hline
   p$_x$             & 0.22     &  0.22     &   0.77     & 0.75   &  0.67  &  0.72     &  0.20 &  0.17    \\
   p$_z$             & 0.22     &  0.22     &   0.79     & 0.75   &  0.74  &  0.74     &  0.20 &  0.15   \\     
   d$_{xy}$          &          &           &            &        &        &           &  0.88 &  0.12   \\
   d$_{xz}$          &          &           &            &        &        &           &  0.95 &  0.08   \\
   d$_{z^2}$         &          &           &            &        &        &           &  0.98 &  0.16   \\ \hline\hline
\end{tabular}
\end{table}

\section{Conclusions}
\label{concl}

In summary, we have compared the electronic structure of ferromagnetic
(Ga,Mn)As and (Ga,Mn)N within standard LSDA and within our 
pseudo-SIC approach which sets the KS potential free of the spurious 
self-interaction part. Our results indicate two main conclusions. 
First, that the LSDA picture of (Ga,Mn)As as a half-metal, with the
Fermi energy crossing three weakly correlated bands which derive from 
Mn d - As p hybridized states, is indeed a good description, largely
unchanged when the self-interaction is corrected. However for (Ga,Mn)N 
the LSDA description is inadequate because the d electrons are strongly 
correlated, and thus the pseudo-SIC description is markedly different 
from that of the LSDA. 
Second, that the behavior of (Ga,Mn)N is qualitatively different
from that of (Ga,Mn)As. Three flat bands of d$_{xy}$, d$_{x^2-y^2}$, and
d$_{z^2}$ character are partially occupied by two spin-polarized 
electrons, confined 
within the Mn-centered tetrahedron, thus effectively describing 
a highly confined 
hole around the Fermi energy which may be more consistent with the 
formation of a Zhang-Rice polaron, rather than with the free-hole mediated 
double-exchange mechanism of the Zener model.
   
S.S. thanks Enterprise Ireland (grant SC-2002/10) for financial support.
A.F. and N.A.S.'s development of the pseudo-SIC method has been supported 
by the NSF under grant number DMR-0312407, and DMSs work by the ONR under
grant number N00014-00-1-0557.



\begin{thebibliography}{99}
\bibitem{note}
 Notice that the difference between the electronic properties of 
 WZ (Ga,Mn)N and ZB (Ga,Mn)As has a chemical origin, whereas structural 
 differences have little or no importance. This is definitely true for 
 bulk GaN and GaAs. Also, other LSDA calculations comparing WZ and ZB 
 (Ga,Mn)N found little differences in their respective DOS.
 See Ref.\onlinecite{kalotov} and \onlinecite{fong}.
\end{thebibliography}
\end{document}